\documentclass[10pt]{article}
\usepackage{cite}
\usepackage{amsmath,amssymb,amsfonts}
\usepackage{algorithmic}
\usepackage{graphicx}
\usepackage{amsthm,bm,booktabs,hyperref,xcolor, cite}

\usepackage{textcomp}
\def\BibTeX{{\rm B\kern-.05em{\sc i\kern-.025em b}\kern-.08em
    T\kern-.1667em\lower.7ex\hbox{E}\kern-.125emX}}
    
\newtheorem{proposition}{Proposition}
\newcommand{\dn}{\partial_n}
\newcommand{\ve}{\varepsilon}
\newcommand{\rr}{\mathbf r}
\newcommand{\nn}{\hat{\mathbf n}}
\newcommand{\tth}{\hat{\boldsymbol{\theta}}}
\newcommand{\zz}{\hat{\mathbf z}}
\newcommand{\kob}{k_{\rm obj}}
\newcommand{\Lcut}{L_{\rm cut}}
\newcommand{\Lon}{L_{\rm on}}

\usepackage[margin=1in]{geometry}
    
\begin{document}
\title{Windowed Singularity Subtraction for the Spectral Integral Method}
\author{Ergun Simsek \\
\small{Department of Computer Science and Electrical Engineering} \\
\small{ University of Maryland Baltimore County} \\
\small{ 1000 Hilltop Circle, Baltimore, MD 21250, USA} \\
\small{E-mail: simsek@umbc.edu} }

\maketitle

\begin{abstract}
In 2004, the spectral integral method (SIM) was proposed to solve TM$_z$ scattering by homogeneous two-dimensional dielectric objects and perfect electrical conductors with smooth boundaries. Its error decreases exponentially as the number of boundary points increases linearly. Plasmonic resonances, however, require TE$_z$ polarization and permittivities with a negative real part. The extension of SIM to this case shows that the original global singularity subtraction suffers catastrophic cancellation, with a round-off floor that grows exponentially with the attenuation accumulated across the object. To resolve this issue, this work introduces a windowed singularity subtraction that limits this cancellation amplification to an exponential factor that the user can set independently of the object size, provided the window is adequately resolved. Numerical results verify the accuracy of the proposed method.
\end{abstract}

\section{Introduction}
Boundary integral equations discretized with global trigonometric bases achieve exponential convergence for smooth, analytic boundaries, provided the logarithmic singularity of the two-dimensional (2D) Green's function is treated analytically \cite{kress91,hu95,trefethen14}. The spectral integral method (SIM) uses this idea for TM$_z$ scattering by homogeneous dielectric and perfectly electric conductor (PEC) cylinders \cite{sim01}, where the cylinder's cross section lies in the $xy$-plane, and the incident electric field is polarized along the $z$ direction. SIM splits the Green's function into a smooth part and the product of a logarithmic kernel with an entire function, and evaluates both with fast Fourier transforms (FFTs). Later, the SIM was extended to objects embedded in layered media \cite{simLM} and to three dimensions \cite{sim3D}. In all of these works, the scatterers were lossless dielectrics or PEC.

Metallic nanostructures support plasmonic resonances in both polarizations when the field can vary along all three directions, as in chains of metal nanoparticles \cite{spr,lmCDA}. For a cylinder that is invariant along $z$ and long compared with the wavelength, however, a localized surface-plasmon resonance requires TE$_z$ polarization ($\mathbf H^{\rm inc}=H_z\zz$), because only then does the electric field have a component normal to the boundary that induces surface charges. Extending SIM to TE$_z$ is therefore needed to study plasmonic resonances of long metallic scatterers and plasmonic waveguides \cite{Oulton}.

Other high-order integral-equation methods have been developed for electromagnetic scattering by penetrable objects. For inhomogeneous two-dimensional scatterers, the Lippmann--Schwinger equation was solved with truncated Fourier expansions of the Green function, giving a high-order solver that runs in $O(N\log N)$ operations even for discontinuous refractive indices \cite{brunoSei00}, and Bruno and Hyde proved that this Fourier-smoothing approach converges with higher order, with rates determined by the regularity of the scatterer \cite{brunoHyde05}. For homogeneous objects, Balaban \emph{et al.} discretized boundary integral equations, including the second-kind M\"uller equations for arbitrary dielectric cylinders, with Nystr\"otype quadratures adapted to smooth, logarithmic, and hypersingular kernels, and applied them to plasmon-assisted H-polarized scattering by silver nanostrips \cite{balaban12}. Sun \emph{et al.} introduced a field-only, fully desingularized surface integral formulation for three-dimensional dielectric bodies \cite{sun20}. Closest to the present problem, Helsing and Karlsson treated the two-dimensional transmission problem for the axial magnetic field of a cylinder with complex permittivity and sharp corners, combining second-kind Kleinman--Martin equations with Nystr\"om discretization, local product integration, and recursively compressed inverse preconditioning, and obtained highly accurate fields even for real negative permittivities \cite{helsing18}. In these methods, the singular part of the kernel is handled either within a volume formulation or locally, on the boundary elements adjacent to the singularity, as in the partition-of-unity localization of high-order surface solvers \cite{bruno01}. Methods that instead remove the logarithmic singularity globally, as SIM does, face a different difficulty, which has not been addressed so far.

Extending SIM to TE$_z$ scattering requires the careful treatment of two related details. First, for metals in the visible part of the spectrum, the interior wavenumber $\kob$ is complex, so the branches of square roots, logarithms, and Hankel functions must be chosen consistently. Second, when $\operatorname{Im}(\kob)D$ is large, where $D$ is the largest chord of the scatterer, the global singularity subtraction of \cite{sim01} generates exponentially large intermediate terms that cancel catastrophically. This work proposes a windowed singularity subtraction that removes this limitation while preserving the structure of SIM: $N$ unknowns per field, FFT-based kernels, and no special quadrature. The cancellation mechanism and its round-off floor are analyzed, the non-analyticity of any localized splitting and its implications for super-algebraic convergence are established, and a signed-curvature self-term valid for non-convex boundaries is provided. An empirically calibrated rule for choosing $N$ is also proposed.

The paper is organized as follows. Section~\ref{sec:form} presents a compact formulation of electromagnetic scattering that accommodates both TE$_z$ and TM$_z$ excitations, introduces the windowed singularity subtraction, and discusses its convergence properties. Section~\ref{sec:val} validates the method against the Mie series for circular cylinders and an FDTD solver for a non-convex cylinder, investigates the window parameters and conditioning, evaluates the proposed rule for choosing $N$, reports the computational cost, and then we conclude.

\section{Formulation}\label{sec:form}
A homogeneous object with relative permittivity $\ve_{\rm obj}$ and boundary $\partial\Omega$ is embedded in a lossless background $\ve_b>0$. Both media are nonmagnetic, and the time dependence is $e^{-i\omega t}$. The scatterer is passive, $\operatorname{Im}\{\ve_{\rm obj}\}\ge0$, and $\operatorname{Re}\{|\ve_{\rm obj}|\}>0$. The wavenumbers inside the background and object are $k_b=k_0\sqrt{\ve_b}$ and $\kob=k_0\sqrt{\ve_{\rm obj}}$, respectively, with the principal square root. The unit incident wave is
\begin{equation}
U^{\rm inc}=e^{ik_b(x\cos\theta_{\rm inc}+y\sin\theta_{\rm inc})},
\label{eq:inc}
\end{equation}
so that $\theta_{\rm inc}=0$ corresponds to a propagation along the $+x$ direction.

\begin{figure}[b] 
\centering
\includegraphics[width=0.3\linewidth]{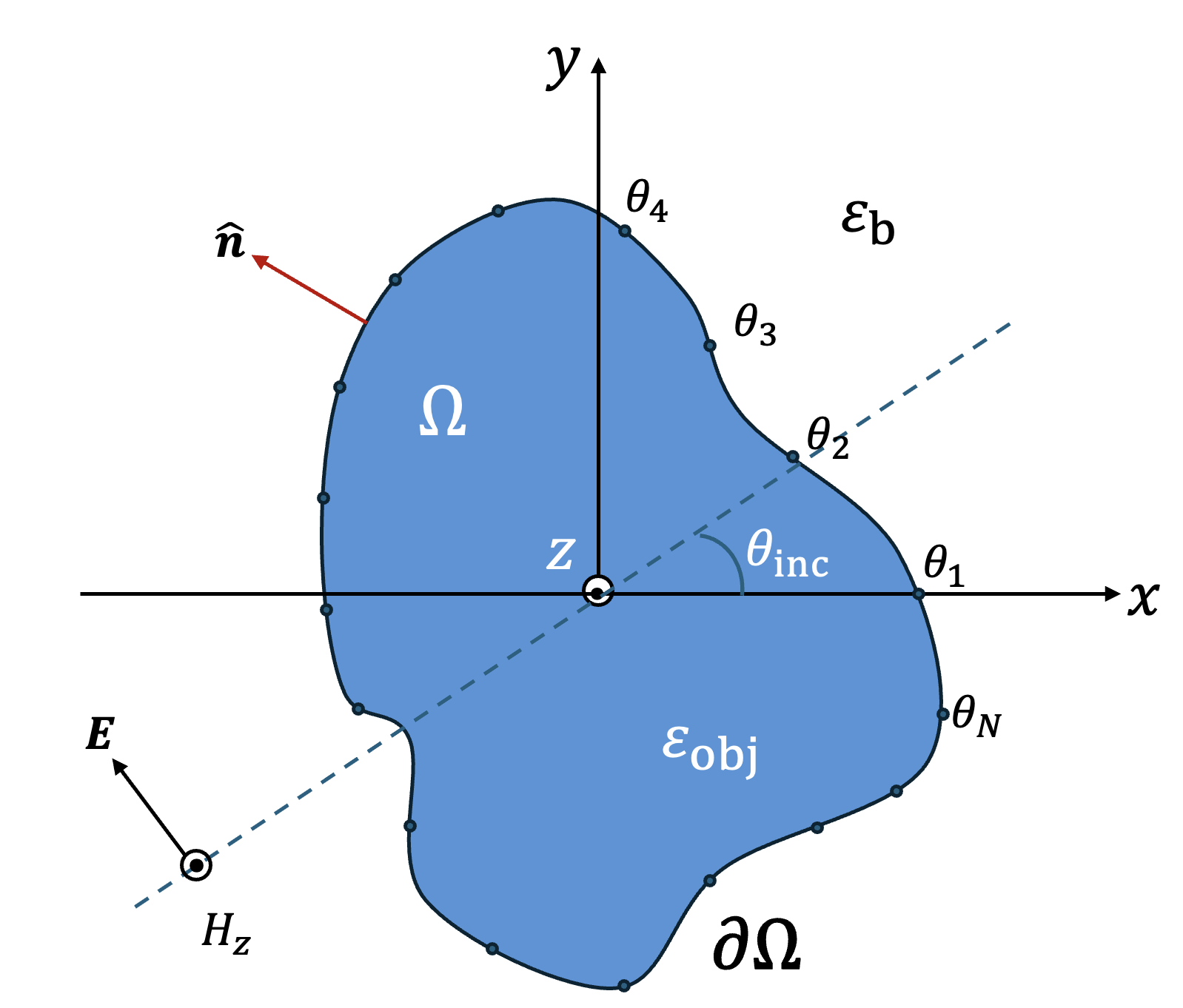}
\caption{A 2D smooth object with relative permittivity $\ve_{\rm obj}$ in a background $\ve_b$. The plane wave propagates in the direction $\theta_{\rm inc}$, measured from the $x$ axis, and $\nn$ is the outward unit normal.}
\label{fig1}
\end{figure}

The boundary is parametrized counter-clockwise, $\rr(\theta)$ with $\theta\in[0,2\pi)$. The Jacobian is $J(\theta)=|\rr'(\theta)|$, the unit tangent is $\tth=\rr'/J$, and the outward unit normal is
\begin{equation}
\nn=\tth\times\zz=\frac{(y'(\theta),\,-x'(\theta))}{J(\theta)} ,
\end{equation}
so that $\zz\times\nn=\tth$ (Fig.~\ref{fig1}). All normal derivatives $\dn=\nn\cdot\nabla$ use this outward normal. 

For TE$_z$ polarization, $\mathbf H=H_z(x,y)\,\zz$ and $\mathbf E=\frac{i}{\omega\ve_0\ve}\nabla H_z\times\zz$. Maxwell's equations can be reduced to $\nabla^2H_z+k_j^2H_z=0$, $j\in\{b,{\rm obj}\}$, in each homogeneous region. No surface currents are present, so the tangential magnetic field is continuous, $H_z^{+}=H_z^{-}$ on $\partial\Omega$, where $+$ and $-$ denote the exterior and interior sides. The tangential electric field is
\begin{equation}
E_t=\tth\cdot\mathbf E=\frac{i}{\omega\ve_0\ve}\nabla H_z\cdot(\zz\times\tth)=-\frac{i}{\omega\ve_0\ve}\,\dn H_z ,
\label{eq:Et}
\end{equation}
Continuity of $E_t$ gives $\dn H_z^{-}=\rho\,\dn H_z^{+}$, where $\rho={\ve_{\rm obj}}/{\ve_b}$. For metals, $\operatorname{Re}\{\rho\}<0$ and the normal derivative changes sign across the interface. This sign change supports surface plasmons, which, for a geometry invariant along $z$, exist only under TE$_z$ excitation.

Let $U$ denote $H_z$ for TE$_z$ or $E_z$ for TM$_z$. The derivative of the Green's function, $G_j(\rr,\rr')=\frac{i}{4}H^{(1)}_0(k_jR)$ with $R=|\rr-\rr'|$, is then
\begin{equation}
\frac{\partial G_j}{\partial n'}=-\frac{ik_j}{4}H^{(1)}_1(k_jR)\,\frac{(\rr'-\rr)\cdot\nn'}{R},
\label{eq:dG}
\end{equation}
where $H^{(1)}_0(\cdot)$ and $H^{(1)}_1(\cdot)$ are Hankel functions of the first kind, of orders zero and one, respectively. For $\rr\notin\bar\Omega$, Green's representation of the radiating field $U-U^{\rm inc}$ in the exterior, with the background Green's function $G_b$ and the outward normal, is
\begin{equation}
U-U^{\rm inc}=\oint_{\partial\Omega}\Big[(U-U^{\rm inc})\frac{\partial G_b}{\partial n'}-G_b\,\dn\big(U^{+}-U^{\rm inc}\big)\Big]ds'.
\end{equation}
The incident field is regular in $\Omega$ and satisfies the background Helmholtz equation there, so Green's second identity in $\Omega$ with the same $G_b$ gives, for $\rr\notin\bar\Omega$,
\begin{equation}
0=\oint_{\partial\Omega}\Big[U^{\rm inc}\frac{\partial G_b}{\partial n'}-G_b\,\dn U^{\rm inc}\Big]ds'.
\end{equation}
Adding the two equations removes $U^{\rm inc}$ from the integrand:
\begin{equation}
U(\rr)-U^{\rm inc}(\rr)=\oint_{\partial\Omega}\Big[U\,\frac{\partial G_b}{\partial n'}-G_b\,\dn U^{+}\Big]ds',\  \rr\notin\bar\Omega .
\label{eq:extrep}
\end{equation}
In the interior, Green's representation of the total field $U$, which satisfies the interior Helmholtz equation, uses the interior Green's function $G_{\rm obj}$:
\begin{equation}
U(\rr)=\oint_{\partial\Omega}\Big[G_{\rm obj}\,\dn U^{-}-U\,\frac{\partial G_{\rm obj}}{\partial n'}\Big]ds',\qquad \rr\in\Omega ;
\label{eq:intrep}
\end{equation}
As $\rr$ approaches $\partial\Omega$, the single-layer potentials are continuous, and the double-layer potential $\oint U\,\partial_{n'}G\,ds'$ tends to its principal value $U/2$ from outside and $-U/2$ from inside. Equations~\eqref{eq:extrep} and \eqref{eq:intrep} therefore give
\begin{align}
\tfrac12U(\rr)&=U^{\rm inc}(\rr)+\oint_{\partial\Omega}\Big[U\,\frac{\partial G_b}{\partial n'}-G_b\,\dn U^{+}\Big]ds',
\label{eq:ext}\\
\tfrac12U(\rr)&=\oint_{\partial\Omega}\Big[G_{\rm obj}\,\dn U^{-}-U\,\frac{\partial G_{\rm obj}}{\partial n'}\Big]ds',
\label{eq:int}
\end{align}
for $\rr\in\partial\Omega$. With the unknowns $\varphi=U|_{\partial\Omega}$ and $\psi=\dn U^+$, and with $\dn U^-=\rho\psi$ from the continuity equation, we obtain
\begin{align}
\tfrac12\varphi-\oint\varphi\,\partial_{n'}G_b\,ds'+\oint G_b\,\psi\,ds'&=U^{\rm inc},\label{eq:extTE}\\
\tfrac12\varphi+\oint\varphi\,\partial_{n'}G_{\rm obj}\,ds'-\rho\oint G_{\rm obj}\,\psi\,ds'&=0 .
\label{eq:intTE}
\end{align}
In TE$_z$, $\varphi=H_z$ and $\psi=\dn H_z=i\omega\ve_0\ve_b\,E_t$ are, up to sign and constant factors, the equivalent electric and magnetic surface currents $\mathbf J_s=\nn\times\mathbf H$ and $\mathbf M_s=-\nn\times\mathbf E$; in TM$_z$ the roles are interchanged. For $\rho=1$ these are the TM$_z$ equations of \cite{sim01}.  

On $N$ equispaced nodes $\theta_m=2\pi(1)/N$ and $m=1,2, \ldots, N$, the unknowns are expanded as truncated Fourier series, $\varphi(\theta)=\sum_n\varphi_ne^{in\theta}$ and $\psi(\theta)=\sum_n\psi_ne^{in\theta}$. The small-argument expansions (9.1.13 and 9.1.11 of \cite{abramowitz}; 10.8.2 and 10.8.1 of \cite{dlmf})
\begin{align}
Y_0(z)&=\tfrac{2}{\pi}\big[\ln\tfrac z2+\gamma\big]J_0(z)+\tfrac{2}{\pi}\sum_{k\ge1}(-1)^{k+1}h_k\frac{(z^2/4)^k}{(k!)^2},\label{eq:Y0}\\
Y_1(z)&=\tfrac{2}{\pi}\ln\tfrac z2\,J_1(z)-\tfrac{2}{\pi z}\notag\\
&\quad-\tfrac1\pi\sum_{k\ge0}(-1)^k\big[h_k+h_{k+1}-2\gamma\big]\frac{(z/2)^{2k+1}}{k!\,(k+1)!},\label{eq:Y1}
\end{align}
where $h_0=0$, $h_k=1+\frac12+\dots+\frac1k$ is the harmonic number, and $\gamma$ is Euler's constant (the bracket equals $\psi(k+1)+\psi(k+2)$ in terms of the digamma function), show that each Bessel function of the second kind consists of the logarithm times an entire function, $\ln(z/2)J_\nu(z)$, plus a remainder. The remainder is entire for $Y_0$ and has a simple pole for $Y_1$. The pole of $Y_1$ is canceled by geometry (because the boundary is smooth). On a smooth curve $(\rr'-\rr)\cdot\nn'=\tfrac12\kappa R^2+O(R^3)$, where
\begin{equation}
\kappa=\frac{x'y''-y'x''}{J^3}
\label{eq:kappa}
\end{equation}
is the signed curvature (positive for a convex boundary traversed counter-clockwise). Only the logarithm is genuinely non-smooth. Since $\ln R=\ln|2\sin[(\theta-\theta')/2]|+\text{smooth}$, the kernels are split as
\begin{align}
\bar G_j&=G_j+\frac{1}{2\pi}\ln\Big|2\sin\frac{\theta-\theta'}{2}\Big|\,J_0(k_jR)\,\chi_j(\theta-\theta'),\label{eq:Gbar}\\
\bar H_j&=\frac{\partial G_j}{\partial n'}-\frac{k_j}{2\pi}\ln\Big|2\sin\frac{\theta-\theta'}{2}\Big|\notag\\
&\qquad\times J_1(k_jR)\,\frac{(\rr'-\rr)\cdot\nn'}{R}\,\chi_j(\theta-\theta'),\label{eq:Hbar}
\end{align}
where $\chi_j\equiv1$ in \cite{sim01} and $\chi_{\rm obj}$ is the window introduced below. The logarithmic parts are discrete convolutions with the known Fourier coefficients of $\ln|2\sin(t/2)|$, namely $a_0=0$ and $a_n=-1/(2|n|)$, and are evaluated with zero-padded FFTs.


\noindent The principal logarithm is applied only to $\kob R/2$ in the kernels and to $\kob J/2$ in the self-term, Eq. \eqref{eq:selfG}. The lossless limits are covered: a lossless dielectric ($\ve_{\rm obj}>0$) gives $\arg\kob=0$, and a lossless metal ($\ve_{\rm obj}<0$) gives $\arg\kob=\pi/2$, both inside the admissible range. The epsilon-near-zero limit ($\ve_{\rm obj}\to0$) is excluded: there $\kob\to0$, the self-term $\ln(\kob J/2)$ diverges, and the interior equation degenerates into a Laplace problem that requires a separate limiting treatment. Active media ($\operatorname{Im}\ve_{\rm obj}<0$) and lossy backgrounds are also outside the scope of this work. 

As $\theta'\to \theta$, $R=J|\theta-\theta'|+O(|\theta-\theta'|^2)$ and $\ln|2\sin[(\theta-\theta')/2]|=\ln|\theta-\theta'|+O(|\theta-\theta'|^2)$. With $G=\frac i4J_0-\frac14Y_0$ and Eq.~\eqref{eq:Y0},
\begin{equation}
\bar G_j(\theta,\theta)=\frac{i}{4}-\frac{1}{2\pi}\Big[\gamma+\ln\frac{k_jJ(\theta)}{2}\Big].
\label{eq:selfG}
\end{equation}
With $H_1^{(1)}(z)=-2i/(\pi z)+O(z\ln z)$ in Eq.~\eqref{eq:dG},
\begin{equation}
\frac{\partial G_j}{\partial n'}\simeq-\frac{(\rr'-\rr)\cdot\nn'}{2\pi R^2}\;\to\;\bar H_j(t,t)=-\frac{\kappa(\theta)}{4\pi},
\label{eq:selfH}
\end{equation}
because the logarithmic term of Eq.~\eqref{eq:Hbar} is $O(R^2\ln R)$. Equation~\eqref{eq:selfH} is independent of $k_j$. With $\kappa$ signed as in Eq.~\eqref{eq:kappa}, it holds on concave arcs. In the implementation, $\kappa$, $J$ and $\nn$ are evaluated analytically from $\rr'(\theta)$ and $\rr''(\theta)$ of the parametrization, not from finite differences of sampled points; the diagonal entry of the double-layer kernel is then simply Eq.~\eqref{eq:selfH} at $\theta=\theta_m$. 

Collocation at $\theta_m$ gives the system of \cite{sim01}, with the interior single-layer block scaled by $\rho$. For each region $j\in\{b,{\rm obj}\}$, testing point $\theta_m$ and Fourier mode $n$, let
\begin{align}
g^j_{mn}&=\frac{1}{2\pi}\int_0^{2\pi}\bar G_j(\theta_m,\theta')\,e^{in\theta'}J(\theta')\,d\theta',\\
u^j_{mn}&=\frac{1}{2\pi}\int_0^{2\pi}\ln\Big|2\sin\frac{\theta_\theta'}{2}\Big|\,p_j(\theta_m,\theta')\,e^{in\theta'}J(\theta')\,d\theta',
\end{align}
and define $h^j_{mn}$ and $v^j_{mn}$ in the same way with $\bar H_j$ and $q_j$, where $p_j=J_0(k_jR)\chi_j$ and $q_j=J_1(k_jR)\frac{(\rr'-\rr)\cdot\nn'}{R}\chi_j$ are the factors multiplying the logarithm in Eqs.~\eqref{eq:Gbar}--\eqref{eq:Hbar}. The smooth integrals $g$ and $h$ are evaluated with the trapezoidal rule, and $u$ and $v$ are discrete convolutions of the Fourier coefficients of $p_jJ$ and $q_jJ$ with $a_n$. Then $\oint\partial_{n'}G_j\,e^{in\theta'}ds'=2\pi h^j_{mn}+k_jv^j_{mn}$ and $\oint G_j\,e^{in\theta'}ds'=2\pi g^j_{mn}-u^j_{mn}$. Multiplying Eq.~\eqref{eq:extTE} by 2 and Eq.~\eqref{eq:intTE} by $-2$ gives
\begin{equation}
\mathbf C^b\bm\varphi+\mathbf D^b\bm\Psi=2\mathbf F,\qquad
\mathbf C^{\rm obj}\bm\varphi+\rho\,\mathbf D^{\rm obj}\bm\Psi=\mathbf 0 ,
\label{eq:system}
\end{equation}
with
\begin{align}
\mathbf C^b&=\mathbf W-2\big(2\pi\mathbf h^b+k_b\mathbf v^b\big),\notag\\
\mathbf C^{\rm obj}&=-\mathbf W-2\big(2\pi\mathbf h^{\rm obj}+\kob\mathbf v^{\rm obj}\big),\\
\mathbf D^j&=2\big(2\pi\mathbf g^j-\mathbf u^j\big),\notag
\end{align}
where $W_{mn}=e^{in\theta_m}$, $F_m=U^{\rm inc}(\rr(\theta_m))$, and $\bm\varphi$ and $\bm\Psi$ are the vectors of Fourier coefficients of $\varphi$ and $\psi$. The $\mathbf D$ blocks have the same form for both regions, and only the sign of the jump term $\mathbf W$ differs. Eliminating $\bm\Psi$ gives the Schur complement
\begin{equation}
\begin{split}
&\mathbf S\bm\varphi\equiv\big[\mathbf C^b-\mathbf D^b(\rho\mathbf D^{\rm obj})^{-1}\mathbf C^{\rm obj}\big]\bm\varphi=2\mathbf F,\\
&\bm\Psi=-(\rho\mathbf D^{\rm obj})^{-1}\mathbf C^{\rm obj}\bm\varphi .
\end{split}
\label{eq:schur}
\end{equation}
For $\operatorname{Re}\{\rho\}<0$, the block $\rho\mathbf D^{\rm obj}$ changes sign relative to the dielectric case, so the two terms of $\mathbf S$ can cancel. This cancellation is the discrete counterpart of a surface-plasmon resonance. In the quasi-static limit of a circular cylinder, the resonance is at $\ve_{\rm obj}=-\ve_b$. For the tested cases, $\operatorname{cond}(\mathbf S)$ peaks there at about $2/\operatorname{Im}\ve_{\rm obj}$ and stays moderate otherwise (Section~\ref{sec:cond}). The windowed subtraction addresses the accuracy of the matrix entries; it does not change this intrinsic sensitivity of the solution near a resonance.

In \cite{sim01}, $\frac{1}{2\pi}\ln|2\sin[(\theta-\theta')/2]|J_0(\kob R)$ is subtracted over the entire boundary. Since $|J_0(\kob R)|\sim e^{\operatorname{Im}\{\kob\}R}/\sqrt{2\pi|\kob|R}$ while $|G_{\rm obj}|\sim e^{-\operatorname{Im}\{\kob\}R}$, the smooth part $\bar G_{\rm obj}$ and the convolution $u_{mn}$ are both of size $e^{\operatorname{Im}\{\kob\}D}$, where $D$ is the largest chord. They cancel to produce an $O(1)$ matrix entry, so the relative error is bounded by
\begin{equation}
\text{error}\;\lesssim\;\epsilon_{\rm mach}\,e^{\operatorname{Im}\{\kob\}D},
\label{eq:roundoff}
\end{equation}
where $\epsilon_{\rm mach}$ is the machine precision. Equation~\eqref{eq:roundoff} estimates the floating-point error of the assembled \emph{matrix entries}. It is a mechanisbased estimate, not a rigorous backward-error bound for the complete algorithm. A relative perturbation of the entries of size $\eta$ changes the computed boundary solution by at most about $\operatorname{cond}(\mathbf S)\,\eta$ in relative terms, and $\operatorname{cond}(\mathbf S)$ is $O(1)$--$O(10^2)$ in our tests (Section~\ref{sec:cond}). The error of the boundary solution additionally contains the discretization error. Equation~\eqref{eq:roundoff} is therefore a floor, not a prediction of the total error, and it describes only the cancellation mechanism that the window controls. The relevant quantity is the attenuation accumulated across the object, $\operatorname{Im}\{\kob\}D=D/\delta$, not the skin depth $\delta$ alone: the global splitting loses about $\operatorname{Im}\{\kob\}D/\ln10$ significant digits. For gold at 800 nm, $\operatorname{Im}\kob\approx0.038$ nm$^{-1}$ (skin depth $\delta=1/\operatorname{Im}\kob\approx26$ nm). A cylinder of diameter $1\,\mu$m gives $e^{38}\epsilon_{\rm mach}\approx7$, so the original method yields huge errors. However, the subtraction only needs to be exact near $\theta=\theta'$. Let $\chi$ be $2\pi$-periodic, with
\begin{equation}
\chi(s)=\begin{cases}\dfrac{f(1-|s|/t_w)}{f(|s|/t_w)+f(1-|s|/t_w)}, & |s|<t_w,\\[2mm] 0, & |s|\ge t_w,\end{cases}
\label{eq:chi}
\end{equation}
where $t_w$ is the half-width of the window, $f(x)=e^{-1/x}$ for $x>0$ and $f(0)=0$. Then $\chi\in C^\infty$ and $\chi(0)=1$, and $1-\chi$ vanishes to all orders at $s=0$, although $\chi$ is not identically equal to one in any neighborhood of $s=0$. The denominator makes $\chi$ and $1-\chi$ a smooth partition of unity on the transition region, and $\chi$ decreases monotonically from 1 at $s=0$ to 0 at $|s|=t_w$. Hence $(1-\chi)\ln|2\sin(s/2)|J_0$ is $C^\infty$ (see the Supplementary Materials and \cite{kress14,Gevrey,rodino93,johnson15,boyd01}), Eqs.~\eqref{eq:Gbar}--\eqref{eq:Hbar} remain smooth, and the FFT part is unchanged. Outside the window, $G_{\rm obj}$ is evaluated directly and is exponentially small. The half-width, measured in the parameter $\theta$, is
\begin{equation}
t_w=\frac{\Lcut}{\operatorname{Im}\{\kob\}\max_tJ(\theta)},
\label{eq:tw}
\end{equation}
where the dimensionless parameter $\Lcut$ bounds the attenuation exponent inside the window: there $R\le t_w\max_tJ$, so $\operatorname{Im}\{\kob\}R\le\Lcut$ and $|J_0(\kob R)|\le e^{\Lcut}$. The cancellation amplification inside the window is therefore at most about $e^{\Lcut}$, and the window extends about $\Lcut$ skin depths along the boundary on each side of the source point. In Eq.~\eqref{eq:chi}, $s$ is the parameter difference $\theta-\theta'$ reduced to $(-\pi,\pi]$, so that the window also covers source points across the parameter seam $\theta=0\equiv2\pi$; the implementation evaluates $\chi$ at the wrapped difference. Because the window is used only when $\operatorname{Im}\{\kob\}D>\Lon$, and every chord satisfies $D\le|\partial\Omega|/2\le\pi\max_tJ$, Eq.~\eqref{eq:tw} gives $t_w<\pi\Lcut/\Lon$. For $\Lcut<\Lon$, the support of $\chi$ therefore lies inside one period, and its periodic extension is $C^\infty$; with the default values, $t_w<1.7$\,rad. The window is applied to the interior kernel only, since the background is lossless, and only if $\operatorname{Im}\{\kob\}D>\Lon$, with the dimensionless switch-on threshold $\Lon$. Below $\Lon$, the round-off floor of the global splitting is acceptable, and its exponential convergence is retained. \footnote{For a target accuracy $\tau$, the estimate~\eqref{eq:roundoff} suggests $\Lon\approx\ln(\tau/\epsilon_{\rm mach})$, which is 13.0 for $\tau=10^{-10}$. The estimate is conservative: for $\operatorname{Im}\{\kob\}D\ge13$ the measured error of the global splitting lies 20 to 30 times below it (Section~\ref{sec:mie}), so $\Lon\approx\ln(\tau/\epsilon_{\rm mach})+3$ suffices in practice. We use $\Lon=15$, for which $\epsilon_{\rm mach}e^{15}\approx7\times10^{-10}$ and the measured global error at the switch is $3\times10^{-11}$ (Section~\ref{sec:winpar})}.

Inserting the large-argument form of $G_b$ (2D free-space Green's function of the Helmholtz equation, evaluated with the background wavenumber) into Eq.~\eqref{eq:ext} gives the scattered field $U^{s}$
\begin{equation}
U^{s}=\sqrt{\frac{2}{\pi k_br}}e^{i(k_br-\pi/4)}\Big[T(\theta)+O\big(\tfrac{1}{k_br}\big)\Big],
\end{equation}
and the far-field amplitude $T(\theta)$ 
\begin{equation}
T(\theta)=\frac i4\oint\Big[-ik_b(\hat{\mathbf r}\cdot\nn')\,\varphi-\psi\Big]e^{-ik_b\hat{\mathbf r}\cdot\rr'}ds' .
\label{eq:T}
\end{equation}
The bistatic scattering width and the cross sections per unit length are
\begin{equation}
\sigma(\theta)=\lim_{r\to\infty}2\pi r|U^s|^2=\frac{4}{k_b}|T(\theta)|^2,
\end{equation}
\begin{equation}
C_{\rm sca}=\frac1{2\pi}\int_0^{2\pi}\sigma\,d\theta,\label{eq:Csca}
\end{equation}
\begin{equation}
C_{\rm ext}=-\frac{4}{k_b}\operatorname{Re}T(\theta_{\rm inc}),\notag
\end{equation}
\begin{equation}
C_{\rm abs}=-\frac{1}{k_b}\oint\operatorname{Im}\big(\bar\varphi\,\psi\big)\,ds .
\label{eq:Cabs}
\end{equation}

With the unit incident amplitude of Eq.~\eqref{eq:inc}, the incident intensity is $I_0=\eta_b/2$, with $\eta_b=Z_0/\sqrt{\ve_b}$; the cross sections are the scattered, extinguished, and absorbed powers per unit length divided by $I_0$, and have the dimension of length. All integrals use the trapezoidal rule in $t$. It is always nice to check the residual $|C_{\rm ext}-C_{\rm sca}-C_{\rm abs}|/C_{\rm ext}$ to see whether it is a very small number, but of course, it is \emph{not} a sufficient accuracy check. 

\section{Numerical results}\label{sec:val}
The method is implemented in MATLAB, and the simulations are done on a 2021 MacBook Pro with 32 GB of RAM. The error is calculated against independent references: the Mie series \cite{bh} for circular cylinders, a converged SIM solution for internal convergence, and FDTD for the non-convex case. Lengths are in nm. We consider two types of materials: BK7 glass is a lossless dielectric, and gold (Au) is a lossy conductor with a negative electrical permittivity in the visible part of the electromagnetic spectrum.

The permittivity of BK7 glass is calculated from its Sellmeier formula \cite{rii} ($\ve_r=2.2955$ at 633 nm). For gold, we use the analytic model of Etchegoin \emph{et al.} \cite{etchegoin}, which was fitted to the Johnson--Christy data \cite{jc}.
At 633 nm, this model gives $\ve_{\rm Au}=-11.253+0.813i$, which corresponds to the complex refractive index $\tilde n=\sqrt{\ve_{\rm Au}}=n_{\rm Au}+ik_{\rm Au}$ with $n_{\rm Au}=0.1211$ and $k_{\rm Au}=3.3567$. For the windowing, we set $\Lcut=8$ and $\Lon=15$.

\subsection{Circular cylinders: comparison with the Mie series}\label{sec:mie}
\begin{figure}[!h]
\centering
\includegraphics[width=0.5\linewidth]{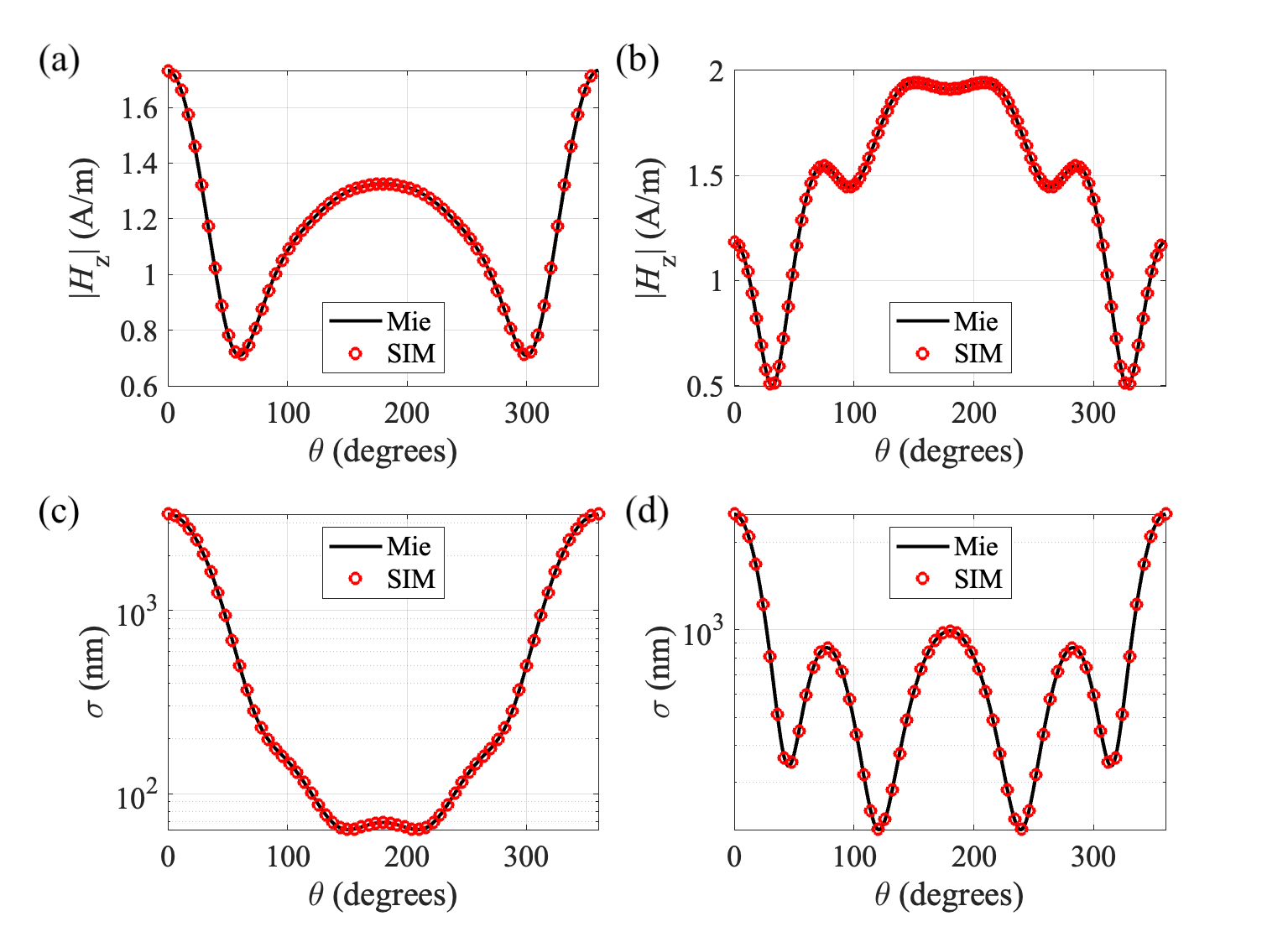}
\caption{BK7 (left, $N=64$) and gold (right, $N=96$) cylinders, $a=200$ nm, $\lambda=633$ nm, $\theta_{\rm inc}=0$ rad. (a,b) $|H_z|$ on the boundary versus polar angle $\theta$. (c,d) Bistatic scattering width. Line: Mie series; circles: SIM.}
\label{figure2}
\end{figure}

Circular cylinders of radius $a=200$ nm are illuminated at $\lambda=633$ nm with $\theta_{\rm inc}=0$ radian and unit amplitude, $H_0=1$\,A/m. For these cases, $\operatorname{Im}\{\kob\}D=13.3$ for gold and $0$ for BK7. Both are below $\Lon$, so the global splitting ($\chi\equiv1$) is used. Figure~\ref{figure2} shows the boundary field and the scattering width. The maximum relative errors with respect to the Mie series are $1.6\times10^{-14}$ (for $|H_z|$) and $2.8\times10^{-15}$ (for $\sigma$) for BK7 with $N=64$, and $6.5\times10^{-12}$ and $8.6\times10^{-12}$ for gold with $N=96$.

\begin{figure}[t]
\centering
\includegraphics[width=0.5\linewidth]{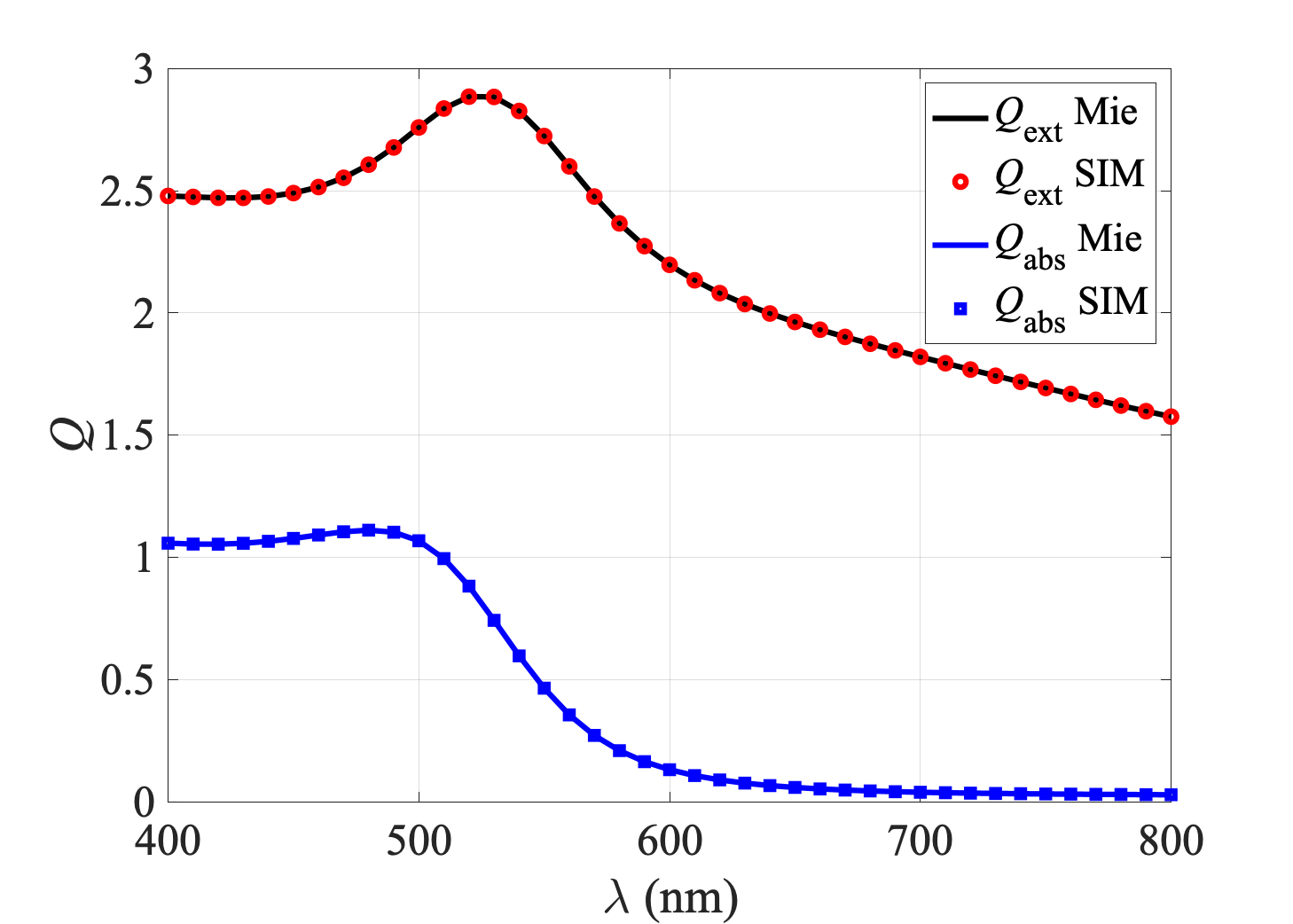}
\caption{Extinction and absorption efficiencies of a gold cylinder ($a=200$ nm, vacuum, TE$_z$). Curves: Mie series; markers: SIM, $N=64$. Maximum deviations: $8.3\times10^{-11}$ ($Q_{\rm ext}$) and $4.8\times10^{-11}$ ($Q_{\rm abs}$).}
\label{figure3}
\end{figure}

Figure~\ref{figure3} shows the extinction and absorption efficiencies, $Q=C/(2a)$, of the gold cylinder in vacuum from 400 to 800 nm. The SIM uses $N=64$. Compared with the Mie series, the maximum deviations are $8.3\times10^{-11}$ in $Q_{\rm ext}$ and $4.8\times10^{-11}$ in $Q_{\rm abs}$, and the energy residual is below $1.2\times10^{-14}$. $Q_{\rm ext}$ peaks at $\approx2.89$ near 520--530 nm. Absorption dominates below about 500 nm, where interband transitions are active, and falls to $Q_{\rm abs}\approx0.03$ at 800 nm.

\begin{figure}[t]
\centering
\includegraphics[width=0.5\linewidth]{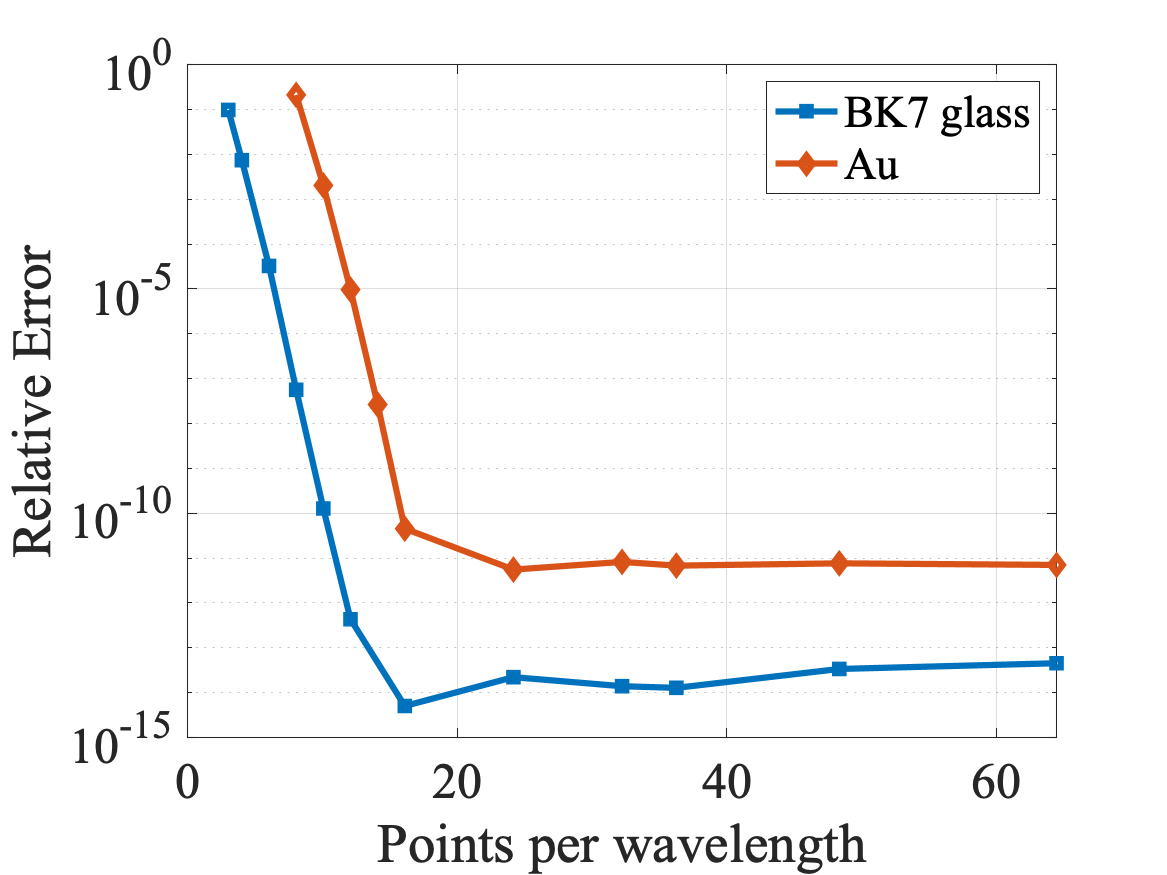}
\caption{Convergence of the TE$_z$ SIM (global splitting) for BK7 and gold cylinders, $a=200$ nm, $\lambda=633$ nm: maximum relative error of the boundary $H_z$ with respect to the Mie series versus the number of boundary points per free-space wavelength.}
\label{figure4}
\end{figure}

Figure~\ref{figure4} shows the maximum relative error of the boundary field, $\max_m|H_z^{\rm SIM}-H_z^{\rm Mie}|/\max_m|H_z^{\rm Mie}|$, against the number of points per wavelength (PPW), $N/(2\pi a/\lambda_0)$, for a BK7 and Au cylinder, where $\lambda_0$ is the free-space wavelength. The error decreases exponentially, as expected for analytic kernels. For BK7 it reaches $10^{-10}$ at 10 PPW and $5\times10^{-15}$ at 16 PPW. For gold, it reaches $3\times10^{-8}$ at 14 PPW and $5\times10^{-11}$ at 16 PPW. Gold needs more points per free-space wavelength because its interior field varies on shorter scales: $|\kob|=3.36\,k_0$ and $\delta\approx30$ nm. Calculated per \emph{interior} wavelength $\lambda/|\tilde n|$, the two materials behave similarly: about 6.7 (BK7) and 4.8 (gold) points give $10^{-10}$. Beyond 16 points per wavelength, the gold error saturates at $\approx7\times10^{-12}$. This plateau is the round-off floor, Eq.~\eqref{eq:roundoff}, of the global splitting: $\epsilon_{\rm mach}e^{13.3}\approx1.3\times10^{-10}$ is an upper bound, and the observed floor lies about 20 times below it.

\begin{figure}[h]
\centering
\includegraphics[width=0.5\linewidth]{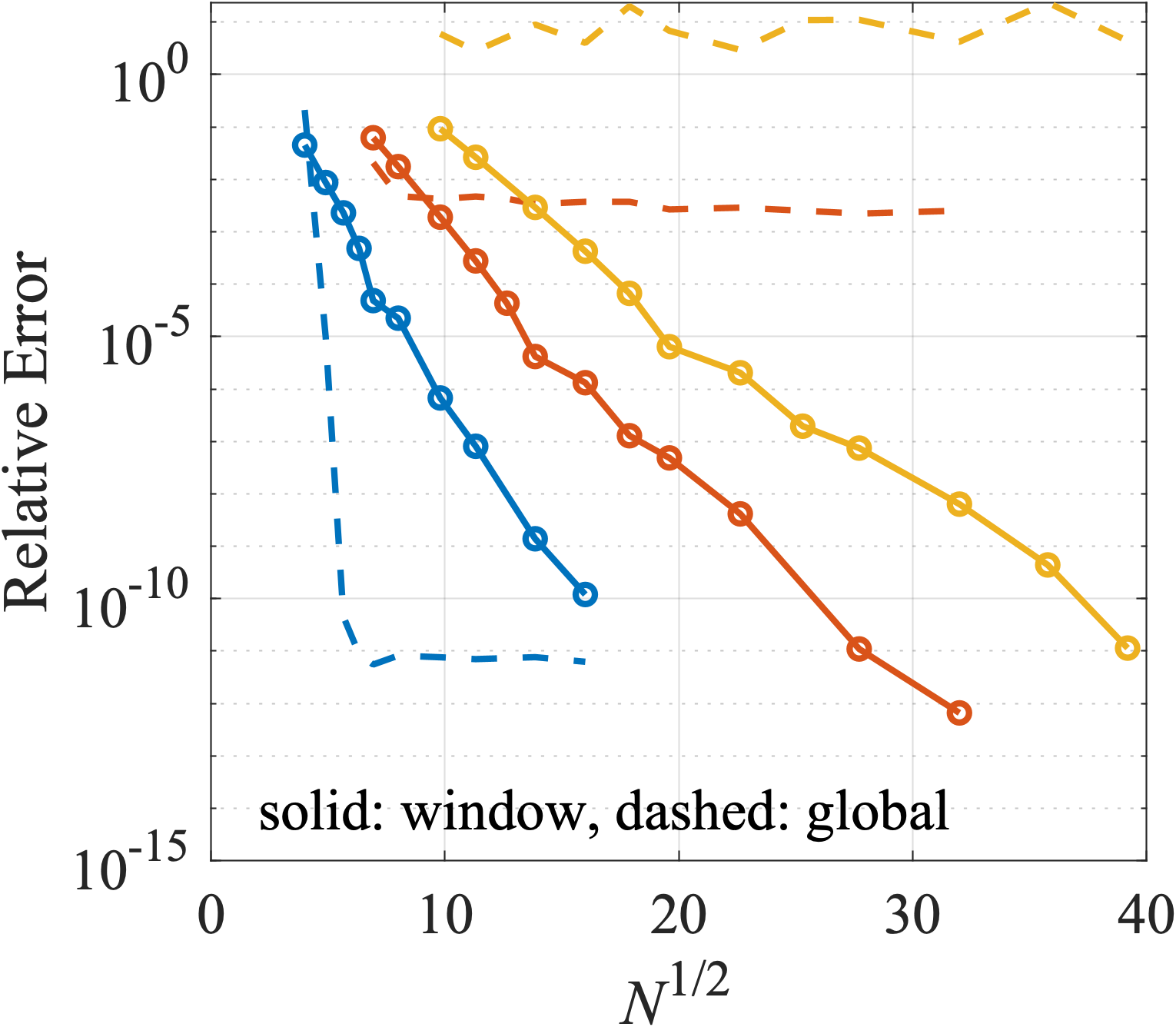}
\caption{Gold cylinders at $\lambda=633$ nm, maximum relative error of the boundary $H_z$ with respect to the Mie series versus $\sqrt N$. Blue: $a=200$ nm ($\operatorname{Im}\{\kob\}D=13.3$); red: $a=500$ nm (33.3); yellow: $a=1000$ nm (66.6). Solid: windowed subtraction ($\Lcut=8$); dashed: global subtraction.}
\label{fig:window1}
\end{figure}

The cases above do not activate the window. To test it, Fig.~\ref{fig:window1} compares the two splittings for gold cylinders with $\operatorname{Im}\{\kob\}D=13.3$, 33.3, and 66.6 ($a=200$, 500, and 1000 nm at $\lambda=633$ nm). In each run, the window is either forced on or forced off. With the global splitting, the error reaches the round-off floor $\approx7\times10^{-12}$ for $a=200$ nm, stalls at $2$--$5\times10^{-3}$ for $a=500$ nm, and is $O(1)$ for $a=1000$ nm, for every $N$. With the window, all three cases converge: the error reaches $1.2\times10^{-10}$ ($a=200$ nm, $N=256$), $7.0\times10^{-13}$ ($a=500$ nm, $N=1024$) and $1.2\times10^{-11}$ ($a=1000$ nm, $N=1536$). The cross sections converge at the same rate; for example, $C_{\rm ext}$ is accurate to $8.7\times10^{-13}$ and $C_{\rm abs}$ to $1.6\times10^{-10}$ for $a=1000$ nm at $N=1536$.

In Fig.~\ref{fig:window1}, the windowed curves are almost straight with increasing negative slopes. We fitted $\ln(\text{error})$ by least squares against $\sqrt N$ and against $N$, using all points with error above $10^{-11}$, so that points dominated by the round-off floor are excluded. The slopes of the fits against $\sqrt N$, which equal $c\sqrt{t_w}$, scale as $\sqrt{t_w}$, so that the coefficient $c$ is nearly the same in the three cases. For the three cases ($t_w=1.20$, 0.480 and 0.240), the error follows the empirical law
$\text{error}\approx C\exp\!\big(-c\sqrt{N t_w}\big)$, where $c=1.52\pm0.01$. 

\subsection{Non-convex boundary: gold rose cylinder}\label{sec:rose}
\begin{figure}[h]
\centering
\includegraphics[width=0.4\linewidth]{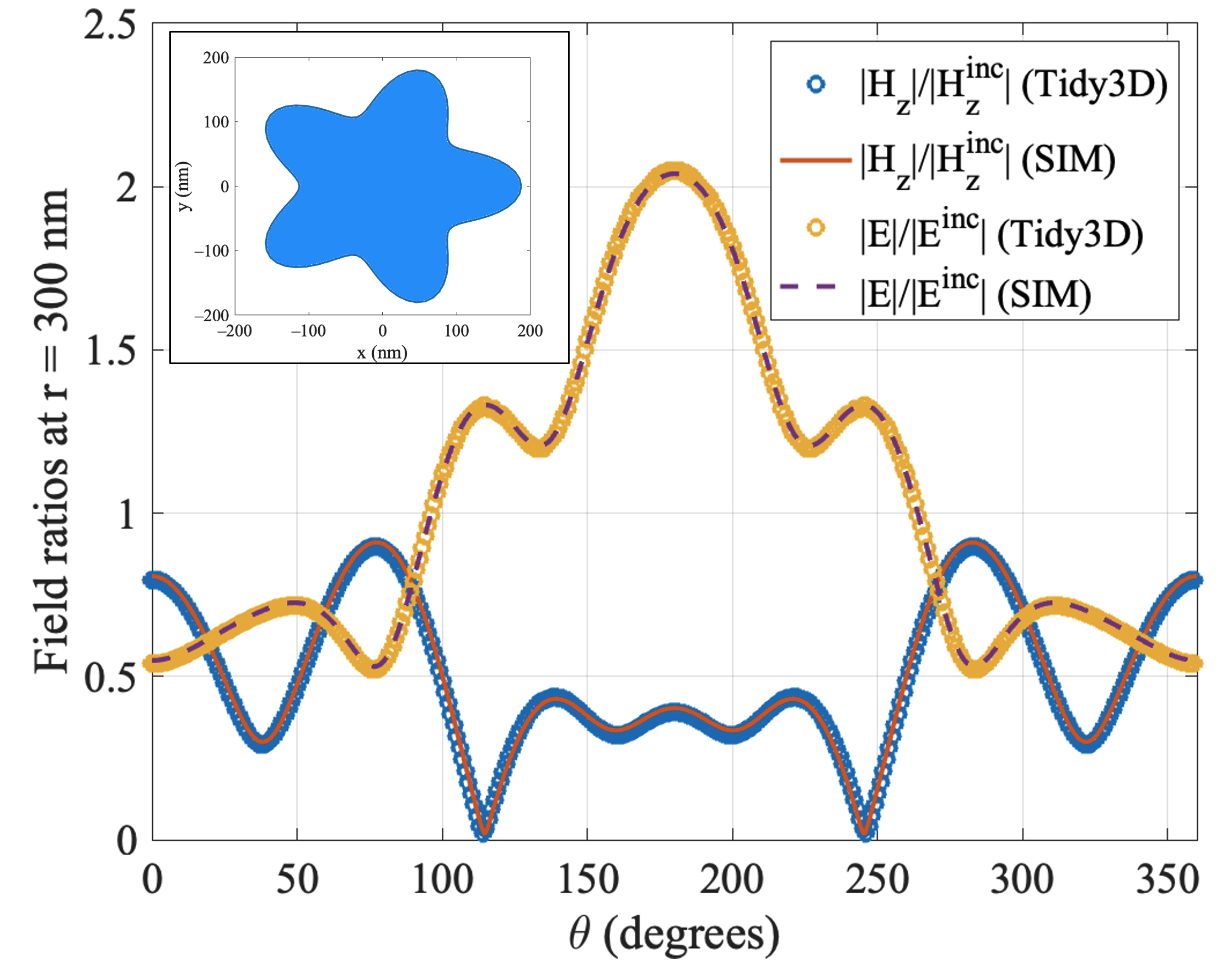}
\caption{Gold rose cylinder [Eq.~\eqref{eq:rose}, $a=150$ nm, $d=0.25$; inset] at $\lambda=633$ nm, TE$_z$ illumination along $+x$ ($\theta_{\rm inc}=0$). Magnitudes of the total fields $|H_z|/|H_z^{\rm inc}|$ and $|\mathbf E|/|\mathbf E^{\rm inc}|$ on a circle of radius $r=300$ nm versus polar angle $\theta$ ($\theta=180^\circ$ is the illuminated side). Curves: SIM ($N=512$). Circles: Tidy3D (uniform 2 nm grid).}
\label{fig:rose}
\end{figure}

For our next numerical study, we chose a scatterer with a non-convex boundary, a five-lobed ``rose'' cylinder defined by
\begin{equation}
r(\theta)=a\,\big[1+d\cos(5t)\big],\qquad t\in[0,2\pi),
\label{eq:rose}
\end{equation}
with $a=150$ nm and $d=0.25$ (inset of Fig.~\ref{fig:rose}). The radius varies between 112.5 and 187.5 nm, and the boundary is concave between neighboring lobes, where the radius of curvature is as small as $15$ nm. The cylinder is gold, in vacuum, at $\lambda=633$ nm, illuminated by a TE$_z$ plane wave and $\theta_{\rm inc}=0$. Here $\operatorname{Im}\{\kob\}D=12.0$, so the global splitting is used. No analytical solution exists, so we compare with a commercial FDTD solver, Tidy3D \cite{tidy3d}. FDTD is not an exact reference, so this comparison establishes agreement at the sub-percent level, not independent verification of the SIM's internal ten-digit convergence.

Relative to a SIM solution with $N=768$, the maximum error of the three cross sections, normalized by $C_{\rm ext}$, is $1.3\times10^{-5}$, $3.1\times10^{-8}$, and $2.1\times10^{-13}$ for $N=96, 160, 256$, respectively. The latter is ten significant digits by \emph{internal} convergence; the independent comparison follows below. The rose needs about four times more points than a circle of the same perimeter because of the high curvature of its necks. The reference values below use $N=512$. Fields outside the cylinder are evaluated from the exterior representation, Eq.~\eqref{eq:ext}, and its gradient, with $\mathbf E=(iZ_0/k_0\ve_b)\,\nabla H_z\times\zz$.

The Tidy3D model is a 2D simulation (zero thickness and periodic boundaries along $z$) on a uniform 2 nm grid, in a $2.0\times2.0\,\mu$m domain terminated by 24-layer perfectly matched layers (PML), with a margin of $0.6\,\mu$m ($\approx1\lambda$) between the scattered-field flux box and the PML. The gold permittivity at 633 nm is represented by a single-pole Lorentz model that reproduces the same $\ve_{\rm obj}$ exactly at this wavelength. The plane wave ($E_y$, $H_z$) is injected with a total-field/scattered-field source. The scattered power is the flux of the scattered field through a closed box outside the total-field region, and the absorbed power is the net inward flux through a box enclosing the cylinder. Both are normalized by the incident intensity from a second simulation without the scatterer. Fields on the probe circle are divided by those of the same empty simulation, which removes the source normalization and phase.

Figure~\ref{fig:rose} compares the magnitudes of the total magnetic and electric fields, normalized by the incident field, on a circle of radius $r=300$ nm. The SIM and FDTD curves are indistinguishable at the scale of the figure over the full angular range. This includes the interference structure on the illuminated side: an electric-field enhancement $|\mathbf E|/|\mathbf E^{\rm inc}|=2.04$ in the backscattering direction, and two near-zeros $|H_z|/|H_z^{\rm inc}|\approx0.02$ at $\theta\approx115^\circ$ and $245^\circ$. Table~\ref{tab:rose} lists the cross sections. The extinction, scattering and absorption cross sections differ by $0.41\%$, $0.35\%$ and $0.05\%$ of $C_{\rm ext}$. Possible sources are the discretization of the FDTD model, since Tidy3D treats metallic media with staircasing by default, so that the 2 nm grid does not represent the curved gold surface exactly while the field varies over the skin depth $\delta\approx30$ nm, and the finite simulation domain and run time.

\begin{table}[t]
\centering
\caption{Cross sections per unit length (nm) of the gold rose cylinder, Eq.~\eqref{eq:rose}, at $\lambda=633$ nm, TE$_z$ illumination along $+x$ ($\theta_{\rm inc}=0$).}
\label{tab:rose}
\small\setlength{\tabcolsep}{4pt}
\begin{tabular}{lccc}
\toprule
 & SIM & Tidy3D & $|\Delta|/C_{\rm ext}$\\
 & ($N=512$) & (2 nm grid) & \\
\midrule
$C_{\rm ext}$ & 1271.23 & 1276.41 & $4.1\times10^{-3}$\\
$C_{\rm sca}$ & 1207.52 & 1212.01 & $3.5\times10^{-3}$\\
$C_{\rm abs}$ & \phantom{1}63.71 & \phantom{1}64.40 & $5.4\times10^{-4}$\\
\bottomrule
\end{tabular}
\end{table}

\subsection{Computational cost}\label{sec:cost}
The assembly requires $O(N^2\log N)$ operations ($N$ rows, each with FFT convolutions of length $2N$), and the dense solve requires $O(N^3)$, so the solve dominates asymptotically. For the sizes considered here, however, the cost is set by the large constant of the assembly, because each matrix entry requires several complex Bessel and Hankel function evaluations, whereas the solve uses optimized dense linear algebra. Table~\ref{tab:time} lists wall-clock times for the complete calculation, including assembly, solve, far field, and the three cross sections, with the windowed singularity subtraction for the gold cylinder with $a=1000$ nm at $\lambda_{\rm inc}=633$ nm. The far field and the cross sections take less than 4\% of the total time for $N\ge256$. Over $128\le N\le1536$, the measured time grows approximately as $N^{1.6}$. This growth is slower than the operation count suggests, because a fixed overhead per solution dominates at small $N$ and vectorized evaluation becomes more efficient as $N$ increases. The local exponent increases with $N$, from about 1.3 at the smallest sizes to 1.8 between $N=512$ and 1536, but it remains below the value of about 2.1 that an $N^2\log N$ cost would give in this range.
\begin{table}[h]
\centering
\caption{Wall-clock time of a complete SIM solution (assembly, solve, far field and cross sections), Au, $a=1000$ nm, $\lambda_{\rm inc}=633$ nm, window on. Medians of seven runs after one warup run.}
\label{tab:time}
\small\setlength{\tabcolsep}{3pt}
\begin{tabular}{lccccccc}
\toprule
$N$ & 128 & 256 & 512 & 768 & 1024 & 1280 & 1536\\
time (s) & 0.06 & 0.16 & 0.46 & 0.88 & 1.48 & 2.13 & 3.43\\
\bottomrule
\end{tabular}
\end{table}

\subsection{Global vs. Windowed Singularity Subtraction}\label{sec:doweneed}
\begin{figure}[h]
\centering
\includegraphics[width=0.4\linewidth]{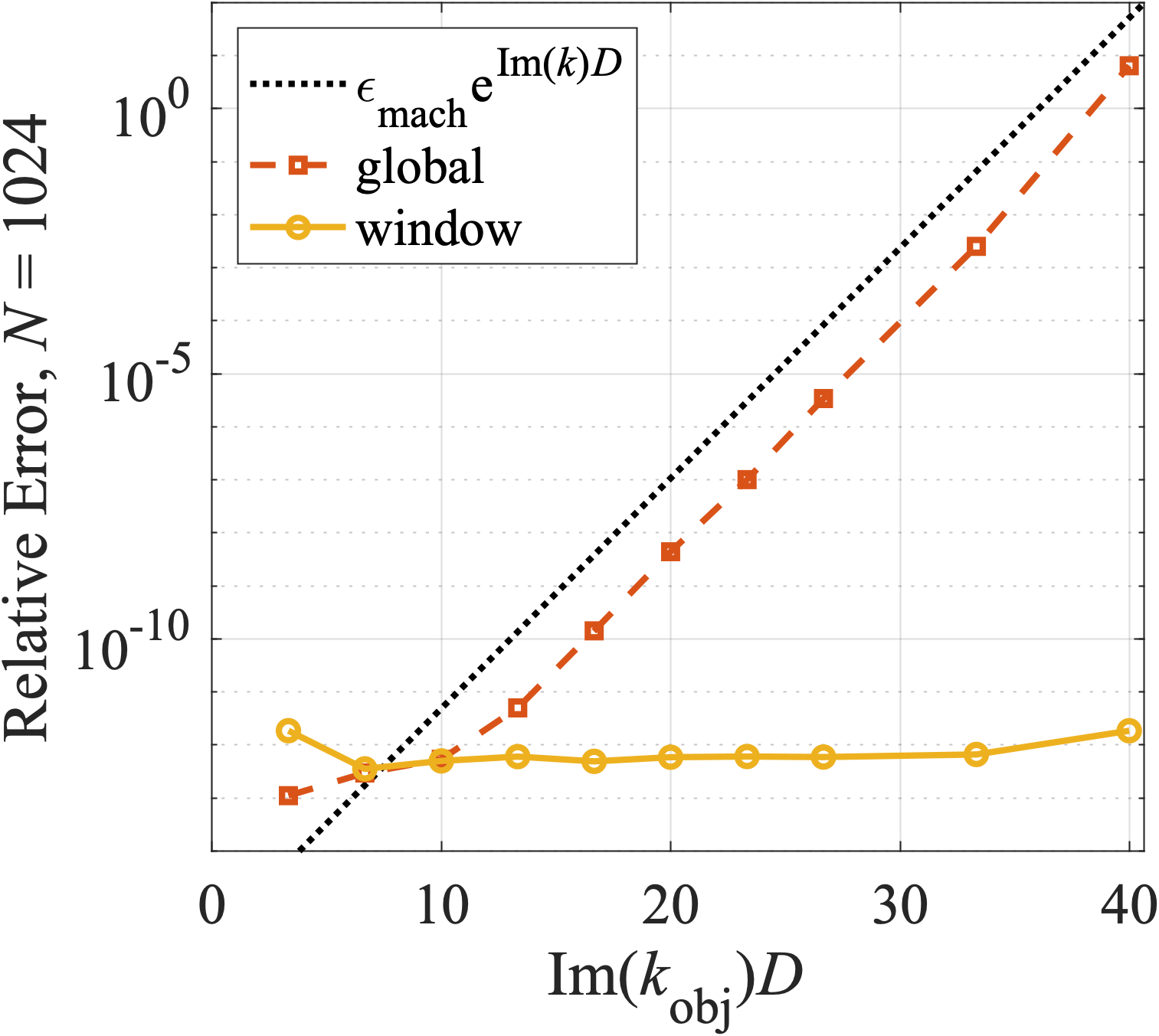}
\caption{Error at $N=1024$ versus $\operatorname{Im}\{\kob\}D$ (gold, $a=50$--600 nm, 633 nm). The global error follows the bound $\epsilon_{\rm mach}e^{\operatorname{Im}\{\kob\}D}$ (dotted), and the windowed error stays at $\le2\times10^{-12}$ for all $\operatorname{Im}\{\kob\}D$ values. }
\label{fig:window2}
\end{figure}

To demonstrate the need for the windowed singularity subtraction, we fix the sampling number to 1024 and calculate the relative error in the total magnetic field with global subtraction and windowed subtraction for circular gold with a radius ranging from 50 nm to 600 nm. At $\lambda=633$ nm, this range corresponds to $\operatorname{Im}\{\kob\}D$ between 3.3 and 40. Figure~\ref{fig:window2} shows that  the global error grows as predicted by Eq.~\eqref{eq:roundoff}, from $10^{-13}$ at $\operatorname{Im}\{\kob\}D=3.3$ to  $2.3\times10^{-3}$ at 33. The windowed error stays at $\le2\times10^{-12}$ for all $\operatorname{Im}\{\kob\}D$ values. For small $a$, however, the global subtraction is more accurate than the windowed subtraction. For small $\operatorname{Im}\{\kob\}D$, the global splitting is limited only by its exponentially convergent discretization error, because its cancellation floor $\epsilon_{\rm mach}e^{\operatorname{Im}\{\kob\}D}$ is negligible; the window removes nothing there but adds a Gevrey-2 transition whose Fourier coefficients decay only root-exponentially, so at equal $N$, it is less accurate.

\subsection{Window parameters}\label{sec:winpar}
\begin{figure}[h]
\centering
\includegraphics[width=0.8\linewidth]{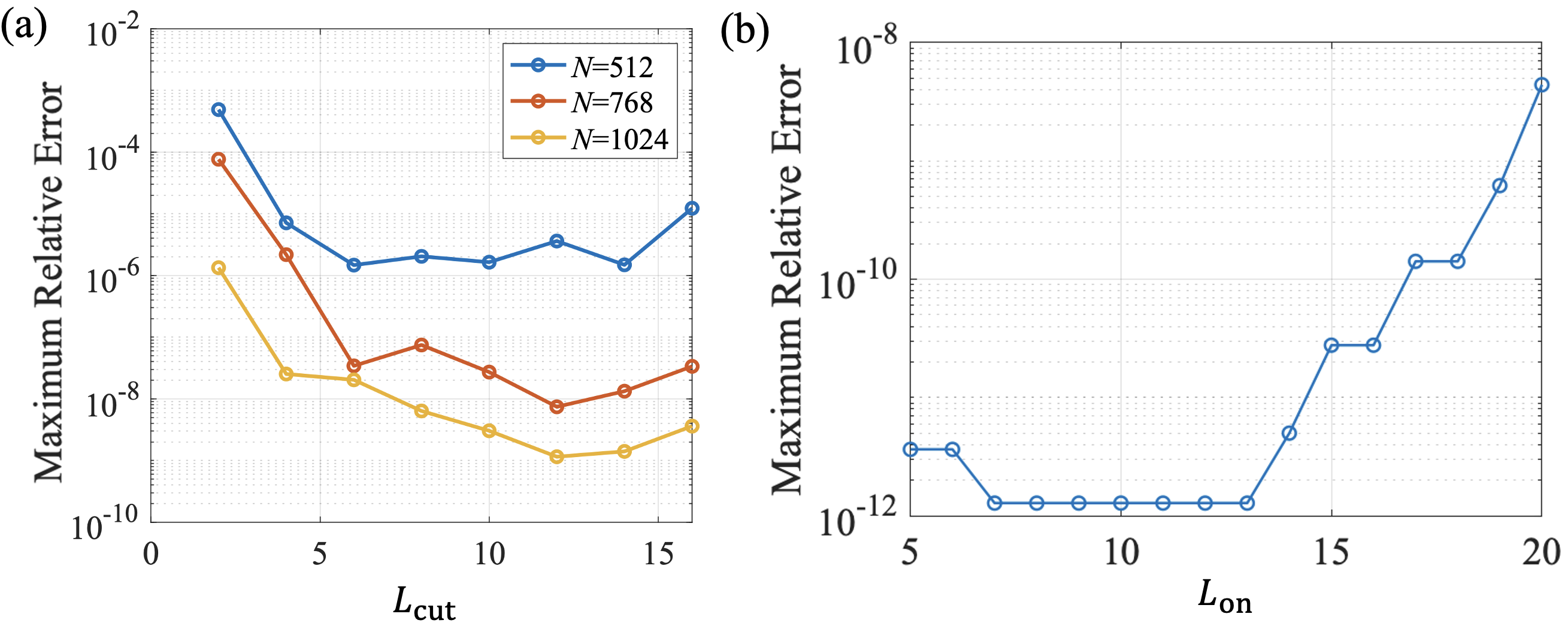}
\caption{(a) Maximum relative error of the boundary $H_z$ with respect to the Mie series versus $\Lcut$ for a gold cylinder, $a=1000$ nm, $\lambda=633$ nm, $N=512$, 768 and 1024. (b) Worst-case maximum relative error of the boundary $H_z$ with respect to the Mie series versus the switch-on threshold $\Lon$, over gold cylinders with $a=50$--600 nm ($\operatorname{Im}\{\kob\}D=3.3$--40) at $\lambda=633$ nm, $N=1024$, $\Lcut=12$.}
\label{fig:window3}
\end{figure}

For the previous examples, $\Lcut$ and $\Lon$ were set to 8 and 15, respectively. Here, we run some additional test cases to investigate their influence on the accuracy. For the former, we assume a gold cylinder with $a=1000$ nm at 633 nm, whose circumference is about $10\lambda$. $L_{\rm on}$ is fixed to 15 and we change $L_{\rm cut}$ from 2 to 16. Figure~\ref{fig:window3} (a) shows how the maximum relative error changes for $2\le\Lcut\le16$ and $N=512$, 768, and 1024. For small $\Lcut$, the window is narrow, its transition is steep, and the error is large: at $\Lcut=2$ it lies between about $10^{-6}$ and $10^{-3}$. It decreases by one to two orders of magnitude up to $\Lcut\approx6$ and then varies by less than one order of magnitude for $8\le\Lcut\le14$, with a shallow minimum near $\Lcut=12$ for $N=768$ and 1024. At $\Lcut=16$ it increases again, consistent with the cancellation factor $e^{\Lcut}$ inside the window, since $\epsilon_{\rm mach}e^{16}\approx2\times10^{-9}$. For the former studies, the default value of $\Lcut$ was 8; apparently, $\Lcut = 12$ would have been a better choice.

The threshold $\Lon$ acts only as a switch: the window is used when $\operatorname{Im}\{\kob\}D>\Lon$ and the global splitting otherwise. Its influence can therefore be seen only over a range of object sizes, not for a single object. We computed gold cylinders at $\lambda=633$ nm with radii $a=50$, 75, \ldots, 600 nm, with $N=1024$ and $\Lcut=12$. For each cylinder, the scattering problem was solved once with the global splitting and once with the windowed splitting. For a given $\Lon$, the error that the method returns is the global result for the radii with $\operatorname{Im}\{\kob\}D\le\Lon$ and the windowed result for the others. Figure~\ref{fig:window3} (b)  shows the largest of these errors over all radii as a function of $\Lon$. For $7\le\Lon\le13$, the worst-case error is constant at $1.3\times10^{-12}$. It is set by the windowed result for the largest cylinder. For $\Lon\ge14$, the global splitting is used for cylinders whose round-off floor is no longer negligible, and the worst-case error grows with $\Lon$. With $\Lcut=12$, any $\Lon$ between 7 and 13 is therefore optimal for this range of sizes. 

\begin{table*}[h]
\centering
\caption{Validation of the sampling rule, Eq.~\eqref{eq:rule}. Dimensions are in nm. $\epsilon$: maximum relative boundary-field error versus Mie or maximum cross-section error relative to $C_{\rm ext}$ versus SIM with $N=1024$ ($N=2048$ for the large ellipse) (non-circular shapes). $N_{10}$: smallest tested $N$ with $\epsilon\le10^{-10}$. The $\epsilon$ column is evaluated at the predicted $N=\max(N_\lambda,N_w,N_g)$.}
\label{tab:extra}
\small\setlength{\tabcolsep}{5pt}
\begin{tabular}{llrrrrcrr}
\toprule
group & case & Im$(\kob)D$ & $N_\lambda$ & $N_w$ & $N_g$ & $\epsilon$ at $N$ & $N_{10}$ & cond$(\mathbf S)$\\
\midrule
shape & Au ellipse $300\times100$, 633 nm & 20.0 & 50 & 394 & 102 & $5.5\!\times\!10^{-11}$ & 394 & 4.4 \\
 & Au rose $d=0.35$, 633 nm & 12.9 & 56 & -- & 470 & $5.3\!\times\!10^{-13}$ & 384 & 11.1\\
 & Au ellipse $600\times200$, 633 nm & 40.0 & 100 & 786 & 102 & $1.5\!\times\!10^{-10}$ & 1024 & 5.9 \\
resonance & Au circle $a=50$, 506 nm& 2.3 & 10 & -- & 16 & $4.3\!\times\!10^{-13}$ & 16 & 2.7\\
 & circle $a=20$, $\ve=-1.02+0.02i$, 633 nm & 0.4 & 2 & -- & 16 & $3.7\!\times\!10^{-14}$ & 16 & 96.6\\
material & circle $a=1000$, $\ve=15.07+0.15i$, 633 nm & 0.4 & 270 & -- & 16 & $4.1\!\times\!10^{-13}$ & 128 & 154\\
 & circle $a=1500$, $\ve=2+3i$, 633 nm & 26.7 & 198 & 524 & 16 & $4.6\!\times\!10^{-10}$ & 768 & 29.6\\
\bottomrule
\end{tabular}
\end{table*}
\begin{table*}[h]
\centering
\caption{Sampling rule Eq.~\eqref{eq:rule} versus observed error at the predicted $N$ ($\lambda$ in nm). $\epsilon_H$: maximum relative boundary-field error versus Mie. $\epsilon_C$: maximum cross-section error relative to $C_{\rm ext}$, versus Mie or versus SIM with $N=768$. }
\label{tab:rule}
\small\setlength{\tabcolsep}{6pt}
\begin{tabular}{lrrrrrcc}
\toprule
case & Im$(\kob)D$ & $N_\lambda$ & $N_w$ & $N_g$ & $N$ & $\epsilon_H$ & $\epsilon_C$\\
\midrule
BK7, $a$=200 nm, $\lambda = 633$ nm & 0 & 22 & -- & 16 & 22 & $7.5\!\times\!10^{-12}$ & $1.2\!\times\!10^{-14}$\\
Au, $a$=200 nm, $\lambda = 633$ nm & 13.3 & 48 & -- & 16 & 48 & $6.6\!\times\!10^{-12}$ & $3.2\!\times\!10^{-12}$\\
Au, $a$=500 nm, $\lambda = 633$ nm & 33.3 & 118 & 656 & 16 & 656 & $1.4\!\times\!10^{-10}$ & $6.7\!\times\!10^{-11}$\\
Au, $a$=500 nm, $\lambda = 800$ nm & 38.3 & 134 & 752 & 16 & 752 & $7.6\!\times\!10^{-11}$ & $4.8\!\times\!10^{-11}$\\
Au, $a$=1000 nm, $\lambda = 633$ nm & 66.6 & 234 & 1310 & 16 & 1310 & $2.6\!\times\!10^{-10}$ & $8.3\!\times\!10^{-11}$\\
BK7 rose, $\lambda = 633$ nm & 0 & 22 & -- & 204 & 204 & -- & $1.2\!\times\!10^{-13}$\\
Au rose, $\lambda = 633$ nm & 12.0 & 48 & -- & 204 & 204 & -- & $1.0\!\times\!10^{-11}$\\
\bottomrule
\end{tabular}
\end{table*}
\subsection{Conditioning}\label{sec:cond}
We also calculated the 2-norm condition numbers of the Schur complement, $\operatorname{cond}(\mathbf S)$, Eq.~\eqref{eq:schur}, and reported them in Table~\ref{tab:extra}. Without the window, the computed $\operatorname{cond}(\mathbf S)$ for $a=1000$ nm fluctuates between $3\times10^3$ and $2\times10^5$, which reflects corrupted matrix entries rather than intrinsic ill-conditioning. To probe the plasmon resonance, we swept $\operatorname{Re}\{\ve_{\rm obj}\}$ from $-3$ to $-0.2$ for a small cylinder ($a=20$ nm, 633 nm, $N=64$) with $\operatorname{Im}\ve_{\rm obj}=10^{-3}$, $10^{-2}$ and $10^{-1}$. $\operatorname{cond}(\mathbf S)$ peaks at $\ve_{\rm obj}=-1.0=-\ve_b$, the quasi-static pole, with values of $2.0\times10^3$, $2.0\times10^2$ and $20.1$, i.e.\ $\approx2/\operatorname{Im}\ve_{\rm obj}$. Even at the sharpest resonance, the error with respect to the Mie series stays below $2.5\times10^{-12}$. For the additional cases of Table~\ref{tab:extra}, the condition number ranges from 2.7 to 154. The near-resonant cylinder with $\ve_{\rm obj}=-1.02+0.02i$ gives 97, again close to $2/\operatorname{Im}\ve_{\rm obj}$. 

\subsection{How to choose $N$ for ultra-high accuracy}\label{sec:rule}
We run several other studies: a gold ellipse with 3:1 aspect ratio, a deeper-neck rose with $d=0.35$ (minimum neck radius of curvature 7.8 nm), and a larger 3:1 gold ellipse with the window on; two near-resonance cases: a gold cylinder with $a=50$ nm at its TE$_z$ extinction peak (506 nm) and a small cylinder with $\ve_{\rm obj}=-1.02+0.02i$ close to the quasi-static pole; two other materials: a high-index, low-loss dielectric representative of silicon at 633 nm ($\ve_{\rm obj}=15.07+0.15i$, $a=1000$ nm, with strong Mie resonances) and a strongly absorbing dielectric with $\operatorname{Re}\ve_{\rm obj}>0$ ($\ve_{\rm obj}=2+3i$, $a=1500$ nm, window on). Their numerical analyses are summarized in Table \ref{tab:rule}. Based on these studies, we recommend choosing $N$ as the largest of the three resolution requirements, rounded up to an even integer,
\begin{equation}
N=\max\big(N_\lambda,\,N_w,\,N_g\big),
\label{eq:rule}
\end{equation}
where
\begin{equation}
N_\lambda=p\,\frac{|\partial\Omega|\,n_{\max}}{\lambda},
\end{equation}
\begin{equation}
N_w=\frac{2\pi n_w\operatorname{Im}\{\kob\}\max_t J}{\Lcut}\ \ (\text{if }\operatorname{Im}\{\kob\}D>\Lon),
\end{equation}
\begin{equation}
N_g=c_g\,|\partial\Omega|\max|\kappa|,
\end{equation}
where $p$ is the number of points per interior wavelength, $n_w$ the number of nodes per window half-width, $c_g$ a curvature coefficient, and $n_{\max}=\max(\sqrt{\ve_b},|\sqrt{\ve_{\rm obj}}|)$. The three terms resolve, respectively, the shortest wavelength, the window, and the boundary curvature; $N_g$ provides about $c_g$ nodes per smallest radius of curvature when the nodes are approximately equispaced in arc length.  To find a good set of constants $(p, n_w, c_g)$, a grid search is done on various cases, and it is observed that the constants $p=7$, $n_w=50$ and $c_g=2.5$ lead to a maximum relative error of $10^{-10}$ or so. Tables~\ref{tab:extra} and \ref{tab:rule} list the $N_\lambda, N_w$ and $N_g$ values obtained using these constants for various cases, along with maximum relative errors. The error at the predicted $N$ ranges from $3.7\times10^{-14}$ to $4.6\times10^{-10}$. Where the rule is off, it is conservative for the curvature- and wavelength-limited cases: the silicon cylinder reaches $10^{-10}$ already at $N=128$ instead of 270, and the deeper-neck rose at $N=384$ instead of 470. It underestimates by 23--32\% for the two large windowed cases, which reach $10^{-10}$ at $N=768$ and 1024 instead of 524 and 786. The small resonant cylinders are resolved at the minimum $N=16$, with errors of $4\times10^{-13}$ and $4\times10^{-14}$. Proximity to a resonance therefore did not increase the required $N$ in these tests, although it does raise $\operatorname{cond}(\mathbf S)$. Because this is an empirical guide, we recommend a doubling test ($N$, $2N$) on a quantity other than the energy residual. In all cases of Table~\ref{tab:extra}, the energy residual at the predicted $N$ is below $10^{-13}$, irrespective of the actual error, which ranges over four orders of magnitude. The same root-exponential law describes the three windowed cases of Table~\ref{tab:extra}, which include two non-circular boundaries and a dielectric with $\operatorname{Re}\{\ve_{\rm obj}\}>0$, with $c=1.48$--1.60. Note that the constants ($p=7$, $n_w=50$ and $c_g=2.5$) were chosen from a small benchmark set, and they should be regarded as an empirically calibrated criterion, not an error guarantee.

\section{Conclusion}\label{sec:concl}
The spectral integral method has been extended from TM$_z$ to TE$_z$ scattering by homogeneous two-dimensional objects with complex permittivity, which makes it applicable to lossy dielectrics and to metals that support surface-plasmon resonances. The extension revealed that the original singularity subtraction, which is accurate for small or lossless objects, loses accuracy rapidly as the object becomes larger or more absorbing, because of numerical cancellation rather than insufficient resolution. The proposed windowed singularity subtraction removes this limitation while keeping the simplicity and FFT-based structure of the method, and it is applied only when the object is large enough to need it. Comparisons with the Mie series show that the method reaches errors below $10^{-10}$ for gold cylinders that are several wavelengths in size, including cases in which the original subtraction returns completely wrong results. For a non-convex gold cylinder, the results agree with an independent FDTD solver to within 0.4\%. An empirically calibrated rule for choosing the number of boundary points is also provided, so that users can obtain this level of accuracy without extensive trial runs and oversampling. 
\section*{Data Availability}
All the code to reproduce the results and figures shared in this manuscript can be found at \url{https://github.com/simsekergun/SIM_TE}.

\bibliographystyle{IEEEtran}
\bibliography{refs_SIM_TE} 

\appendix
\section{Propositions}
\begin{proposition}\label{prop:branch}
With the $e^{-i\omega t}$ convention and $\operatorname{Im}\{\ve_{\rm obj}\}\ge0$, let $\sqrt{\cdot}$ and $\ln$ be principal. Then $\arg\kob\in[0,\pi/2]$ for every value of $\operatorname{Re}\left\{\ve_{\rm obj}\right\}$, and $\arg(\kob R)$, $\arg(\kob J/2)\in[0,\pi/2]$ for all $R,J>0$. Consequently, the expansions \eqref{eq:Y0} and \eqref{eq:Y1} of the manuscript, which hold for $|\arg z|<\pi$, never meet the branch cut, and the outgoing-wave factor $e^{i\kob R}$ of $G_{\rm obj}$ never grows: it decays exponentially when $\operatorname{Im}\kob>0$ and is oscillatory in the lossless dielectric limit.
\end{proposition}

\noindent\emph{Proof.} $\operatorname{Im}\left\{\ve_{\rm obj}\right\} \ge0$ means $\arg\ve_{\rm obj}\in[0,\pi]$, and the principal root halves the argument. Multiplying by $R, J>0$ does not change the argument. 

\begin{proposition}\label{prop:nonanalytic}
Let $w(\theta, \theta')$ be real-analytic in a neighborhood of the diagonal and let $(1-w)\ln|\theta-\theta'|\,J_0(kR)$ be $C^\infty$. Then $w\equiv1$ near the diagonal.
\end{proposition}

\noindent\emph{Proof.} Since $J_0(kR)\to1$, smoothness of $(1-w)\ln|\theta-\theta'|$ requires $1-w$ to vanish to infinite order on the diagonal; otherwise the leading term $c\,(\theta-\theta')^m\ln|\theta-\theta'|$ is not $C^m$. A real-analytic function that vanishes to infinite order at a point is identically zero in a neighborhood. \\

A window that removes the logarithmic singularity and is real-analytic near the diagonal must therefore be identically one there, so a localized window is necessarily non-analytic. The proposition does not fix the convergence rate: windows of different regularity classes give different rates, and the analysis below applies to the Gevrey class-2 window of Eq.~\eqref{eq:chi}. Without the window ($\chi\equiv1$), the logarithmic
singularity remains in the single-layer kernel, even for an analytic boundary. However, after analytically separating the universal logarithmic singularity, the remaining kernel is analytic for an analytic boundary. Fourier-based methods that treat the logarithmic part analytically can therefore achieve exponential convergence, provided the resulting discretization is stable \cite{kress91,trefethen14, Gevrey}. The error then decays as $e^{-cN}$ until it reaches the floor~\eqref{eq:roundoff}. With the window, the regularized kernels are products of $\chi$, $1-\chi$ and $\ln|2\sin[(\theta-\theta')/2]|(1-\chi)$ with functions that are analytic in $(\theta, \theta')$ for an analytic boundary. The individual factors $J_1(kR)$ and $(\rr'-\rr)\cdot\nn'/R$ are not analytic across $\theta = \theta'$, because $R=|\rr-\rr'|$ is not, but the combinations that occur in the kernels are. $J_0(kR)$ and $J_1(kR)/R$ are entire functions of $R^2$, which is analytic in $(\theta, \theta')$; $(\rr'-\rr)\cdot\nn'$ and $(\rr'-\rr)\cdot\nn'/R^2$ are analytic, the latter because both numerator and denominator vanish to second order at $\theta = \theta'$; the regular parts of $Y_0$ and $Y_1$, including the pole term of $Y_1$, are therefore analytic after multiplication by the geometric factor of Eq.~\eqref{eq:dG}; and 
\begin{equation}
\ln(kR/2)-\ln|2\sin[(\theta-\theta')/2]|=\ln\frac k4+\frac12\ln\frac{R^2}{\sin^2((\theta-\theta')/2)} 
\end{equation}
is analytic because the last ratio is analytic and positive. The window built from $f(x)=e^{-1/x}$ belongs to the Gevrey class of order 2 \cite{rodino93}, and Gevrey classes are closed under multiplication by analytic functions. The regularized kernels are therefore Gevrey-2, and their Fourier coefficients decay root-exponentially, as $\exp(-c\sqrt{n\,t_w})$ \cite{rodino93,johnson15,boyd01}. Equations~\eqref{eq:extTE}--\eqref{eq:intTE} couple double-layer (second-kind) and single-layer (first-kind) operators with logarithmic kernels; they do not form a pure second-kind system. For such systems, trigonometric collocation and quadrature methods converge when the discrete system is uniformly stable, and the solution error is then bounded by a constant, related to the norm of the discrete inverse, times the kernel approximation error \cite{kress14}. We assume this stability for the passive materials considered here. It degrades near the quasi-static plasmon pole, $\ve_{\rm obj}\to-\ve_b$ with vanishing loss, where the condition number grows as the loss decreases (Section~\ref{sec:cond}). Under this assumption, the windowed method is super-algebraically convergent, and root-exponential convergence is expected. 


\section{Cross sections and energy balance}\label{app:xs}
For TE$_z$, the field in the background is given by
\begin{equation}
\mathbf E=\frac{i}{\omega\ve_0\ve_b}\nabla H_z\times\zz.
\end{equation}

Since $(\nabla H_z\times\zz)\times\zz=-\nabla H_z$ for in-plane gradients, the time-averaged Poynting vector $\mathbf S=\frac12\operatorname{Re}(\mathbf E\times\mathbf H^*)$ satisfies
\begin{equation}
\mathbf S\cdot\nn=\frac{1}{2\omega\ve_0\ve_b}\operatorname{Im}\big(\bar H_z\,\dn H_z\big).
\label{eq:Sn}
\end{equation}
The incident intensity is $I_0=\frac12\eta_b|H_0|^2$ with $\eta_b=Z_0/\sqrt{\ve_b}$, and $\omega\ve_0\ve_b=k_b\sqrt{\ve_b}/Z_0$.

The absorbed power per unit length is the net inward flux through $\partial\Omega$, evaluated on the exterior side with the total field: $P_{\rm abs}=-\oint\mathbf S\cdot\nn\,ds$. Dividing by $I_0$ with $|H_0|=1$ gives $C_{\rm abs}=-k_b^{-1}\oint\operatorname{Im}(\bar\varphi\psi)\,ds$, which is Eq.~\eqref{eq:Cabs}. By Poynting's theorem in the interior, $P_{\rm abs}=\frac{\omega\ve_0}{2}\int_\Omega\operatorname{Im}\ve_{\rm obj}|\mathbf E|^2dA\ge0$ for passive media with the $e^{-i\omega t}$ convention, so the sign is consistent: $C_{\rm abs}>0$ in all gold computations.

On a circle $r\to\infty$, $\mathbf S_s\cdot\hat{\mathbf r}=\frac{\eta_b}{2}|U^s|^2$. With the asymptotic form of $U^s$, $P_{\rm sca}=\frac{\eta_b}{2}\int_0^{2\pi}\frac{2}{\pi k_b}|T|^2d\theta$, so that 
\begin{equation}
C_{\rm sca}=\frac{2}{\pi k_b}\int_0^{2\pi}|T|^2d\theta=\frac1{2\pi}\int_0^{2\pi}\sigma\,d\theta.
\end{equation}

The extinguished power is minus the flux of the interference term, $\mathbf S_{\rm int}=\frac12\operatorname{Re}(\mathbf E^{\rm inc}\times\mathbf H^{s*}+\mathbf E^s\times\mathbf H^{\rm inc*})$, through a large circle. Using $e^{ik_br\cos(\theta-\theta_i)}$ and evaluating the angular integral by stationary phase at $\theta=\theta_i$, where the forward contribution survives, gives the 2-D optical theorem $C_{\rm ext}=-\frac{4}{k_b}\operatorname{Re}T(\theta_i)$. With the Mie coefficients $a_n$ one has $T(\theta)=\sum_na_ne^{in(\theta-\theta_i)}$, and Eqs.~\eqref{eq:Csca}--\eqref{eq:Cabs} reduce to the familiar 
\begin{equation}
C_{\rm sca}=\frac{4}{k_b}\sum|a_n|^2  \qquad {\rm and } \qquad C_{\rm ext}=-\frac{4}{k_b}\sum\operatorname{Re}a_n. 
\end{equation}
Because $C_{\rm ext}$, $C_{\rm sca}$ and $C_{\rm abs}$ are computed from three different formulas, the balance $C_{\rm ext}=C_{\rm sca}+C_{\rm abs}$ is a consistency check of the convention. In Fig.~\ref{figure3}, $C_{\rm ext}$ and $C_{\rm abs}$ are additionally verified individually against the Mie series.

\section{Implementation notes}\label{app:code}
The method is implemented in MATLAB (version R2024b) on a 2021 Apple MacBook Pro with 32 GB of RAM. The computations use double precision throughout. 
\begin{itemize}
\item  $\rr$, $\rr'$ and $\rr''$ are evaluated analytically from the parametrization on $N$ equispaced nodes, with $N$ even. $J$, $\nn$ and the signed $\kappa$ follow from Eqs.~\eqref{eq:kappa}--\eqref{eq:selfH}.
\item For each testing point $\theta_m$, $R$, $(\rr'-\rr)\cdot\nn'/R$, $\ln|2\sin[(\theta_\theta')/2]|$ (set to 0 at $\theta'=\theta_m$) and $\chi(\theta_\theta')$ are computed. $\chi\equiv1$ for the background and when $\operatorname{Im}(\kob)D\le\Lon$, with $D$ the largest distance between two nodes. This sampled value can only underestimate the largest chord by at most about $2\pi\max_\theta J/N$, and it is exact for a circle with even $N$. It therefore only shifts the switch slightly near $\operatorname{Im}(\kob)D=\Lon$, where both splittings are accurate. $J_0,J_1$ (MATLAB's \texttt{besselj} function) are evaluated only where $\chi>0$, and $H^{(1)}_0,H^{(1)}_1$ (MATLAB's \texttt{besselh} function) everywhere. The self-terms, Eqs.~\eqref{eq:selfG} and \eqref{eq:selfH}, are inserted at $\theta'=\theta_m$, with $p=1$, $q=0$ there. No special treatment of small or large complex arguments is needed beyond the library routines, because the window keeps $|\operatorname{Im}(\kob R)|\le\Lcut$ wherever $J_\nu$ is evaluated.
\item The products of $\ln|2\sin[(t-\theta')/2]|$ with $pJ$ and $qJ$ are evaluated as discrete convolutions with $a_n=-1/(2|n|)$, using zero-padded FFTs of length $2N$.
\item  The two dense systems of Eq.~\eqref{eq:schur} are solved with MATLAB's backslash (LU with partial pivoting), without scaling or preconditioning.
\item The Mie series is truncated at $|n|\le x+4x^{1/3}+15$, with $x=k_ba$.\footnote{W. J. Wiscombe, ``Improved Mie scattering algorithms," \emph{Applied Optics}, vol. 19, no. 9, pp. 1505--1509, 1980.} Internal references for non-circular shapes use $N=768$--2048, as stated in each table.
\end{itemize}
The code and the scripts reproducing all figures and tables are available at \url{https://github.com/simsekergun/SIM_TE}.
     
\end{document}